\documentclass[final,3p]{elsarticle}

\graphicspath{{./}{./article_figures/}}

\usepackage[utf8]{inputenc}
\usepackage[T1]{fontenc}

\usepackage{amsmath}
\usepackage{amssymb}
\usepackage{gensymb}
\usepackage[dvipsnames,svgnames,x11names]{xcolor}

\usepackage{float}
\usepackage{tabularx}
\usepackage{subcaption}
\usepackage{hyperref}
\usepackage[switch, modulo, displaymath, mathlines]{lineno}
\newcommand*\patchAmsMathEnvironmentForLineno[1]{
  \expandafter\let\csname old#1\expandafter\endcsname\csname #1\endcsname
  \expandafter\let\csname oldend#1\expandafter\endcsname\csname end#1\endcsname
  \renewenvironment{#1}
  {\linenomath\csname old#1\endcsname}
  {\csname oldend#1\endcsname\endlinenomath}}
  \newcommand*\patchBothAmsMathEnvironmentsForLineno[1]{
  \patchAmsMathEnvironmentForLineno{#1}
  \patchAmsMathEnvironmentForLineno{#1*}}
  \AtBeginDocument{
  \patchBothAmsMathEnvironmentsForLineno{equation}
  \patchBothAmsMathEnvironmentsForLineno{align}
  \patchBothAmsMathEnvironmentsForLineno{flalign}
  \patchBothAmsMathEnvironmentsForLineno{alignat}
  \patchBothAmsMathEnvironmentsForLineno{gather}
  \patchBothAmsMathEnvironmentsForLineno{multline}
}
\hypersetup{
	colorlinks=true,
	linkcolor=blue,
	filecolor=magenta,      
	urlcolor=cyan,
}

\usepackage{placeins} 

\usepackage{xifthen}
\newcommand{\comment}[2][]{{\color{red}
		\ifthenelse{\isempty{#1}}%
		{}
		{\textit{#1}:~}
		#2}}

\journal{Elsevier}

\begin{document}

\begin{frontmatter}



\title{Efficient data-driven flow modeling for accurate passive scalar advection in submesoscale domains}


\author[riteh]{Karlo Jakac\corref{cor}}
\ead{karlo.jakac@uniri.hr}
\author[riteh]{Luka Lanča}
\ead{luka.lanca@uniri.hr}
\author[cnrm]{Ante Sikirica}
\ead{ante.sikirica@uniri.hr}
\author[riteh]{Stefan Ivić}
\ead{stefan.ivic@uniri.hr}

\affiliation[riteh]{organization={Faculty of Engineering, University of Rijeka},
	addressline={Vukovarska 58}, 
	city={Rijeka},
	postcode={51000}, 
	country={Croatia}}

\affiliation[cnrm]{organization={Center for Advanced Computing and Modelling, University of Rijeka},
	addressline={Radmile Matejčić 2}, 
	city={Rijeka},
	postcode={51000}, 
	country={Croatia}}

\cortext[cor]{Corresponding author}

\begin{abstract}
Knowing the sea surface velocity field is essential for various applications, such as search and rescue operations and oil spill monitoring, where understanding the movement of objects or substances is critical. However, obtaining an accurate approximation of these advection processes is challenging, even with modern measuring equipment, such as high-frequency radar or advanced simulations based on oceanic flow models. Therefore this paper presents a data-driven framework to approximate sea surface velocity from spatially distributed observations, thus enabling efficient probability advection modeling across submesoscale domains. The system approximates transient flows by leveraging quasi-steady flow assumptions. To overcome the limitations of point measurements in capturing domain-wide circulation, the method employs a fusion of two simplified 2D flow models to approximate submesoscale dynamics, enabling complete velocity field reconstruction from scattered data. To ensure reliable flow dynamics, the approach iteratively adjusts boundary conditions in numerical simulations to align the simulated flow with observations. Experimental validation in Kvarner Bay using GPS-tracked drifters confirmed the system’s ability to replace computationally intensive transient simulations by approximating flow fields based on model simplifications. The results demonstrate its efficiency across domains, making it a practical tool for real-world submesoscale applications requiring swift passive scalar advection.

\end{abstract}




\end{frontmatter}


\section{Introduction}

The continuous growth of the marine economy and increased traffic bring about some negative aspects, including a higher frequency of accidents and the potential for environmental disasters. Oil spills and hazardous material pollution are among the most common examples, causing significant economic and ecological damage each year. The time required for recovery varies depending on factors such as the severity of the spill, its composition, and the affected ecosystem, with some cases taking decades for full restoration \cite{elliott2000need}. The financial impact of incidents like oil spills can reach billions of dollars, posing a significant threat to marine habitats by contaminating the food web and polluting vast coastal areas \cite{carson2003contingent}. For instance, the Deepwater Horizon explosion in the Gulf of Mexico caused devastating environmental damage and multiple casualties \cite{mariano2011modeling}. Similarly, the Sanchi oil spill, caused by the collision of the Sanchi oil tanker with a cargo vessel in January 2018, remains one of the most serious and polluting tanker accidents of the 21st century \cite{chen2020marine}. Given the severe consequences of such incidents, the need for effective oil spill emergency preparedness and response has become a global priority to minimize damage, protect lives, and reduce economic losses \cite{garcia2016dynamical}.

In recent decades, the lack of reliable predictions for the spread of pollution has made it challenging to protect ecosystems and the economy from the consequences of environmental disasters. The increasing frequency of such incidents has raised concerns within the scientific community about the urgent need for accurate and reliable models to forecast the progression of maritime disasters, providing crucial decision-making support during emergencies \cite{reed1999oil, kirby2010accidental, wang2010modeling, olascoaga2012forecasting}. Various numerical models are commonly/typically used to  predict the movement of pollutants or their concentration in the water. The results from these transport models are crucial, as they are often the only available approach. However, there are various prediction models with different complexities to calculate pollution movement. The effectiveness of each model depends on its structure, applied methods, and, most importantly, the accuracy of input data (e.g., sea currents, wind, and source location), followed by result interpretation. Given potential input errors, these parameter uncertainties must be carefully considered.

Generally, the primary sources of uncertainty in pollution predictions and Search and Rescue (SAR) operations are the inherent complexities of ocean dynamics \cite{sayol2014lagrangian}. Furthermore, due to the chaotic nature of the ocean, governed by the nonlinear Navier-Stokes equations, means that even small variations in initial conditions can lead to significant deviations in forecast fields over time. Additional uncertainties arise from the drifting dynamics model, as an object’s motion at sea is influenced by the ratio of its surface area exposed to wind and water, with the wind drag coefficient typically determined empirically through trial and error. The accuracy of oceanic environmental data also poses a challenge, as operational ocean and atmospheric forecasting models do not resolve small-scale turbulent motions but instead rely on parameterization, meaning sub-grid-scale information is not directly provided. These factors, combined with incorrect initial conditions during the early stages of an accident, contribute to the inherent uncertainty in ocean drift predictions and further affect forecasting accuracy. As a result, numerous studies have proposed various prediction schemes and developed drift prediction models to address these uncertainties. One of the most widely used approaches is the probabilistic model based on the leeway dynamics model \cite{allan1999review, breivik2008operational}, which helps describe uncertainties in the drifting trajectory of objects. A notable advancement was made by \cite{li2019forecasting}, who developed numerical simulation scenarios that incorporated different current and wind forcing data into the forecasting system and varied oil release times to identify key error sources. Additionally, several studies \cite{qazi2014computing, lorente2016characterizing, paduan2013high} have integrated coastal radar current data for computing ocean surface currents and for possible spread of pollutants.

Still, all prediction methods rely on sea surface velocity measurements from various sources. Due to its cost-effectiveness and reliability, satellite tracking of deployed drifters is one of the most widely used methods. In the past two decades, the use of drifters for monitoring purposes has grown considerably, with numerous deployments in various oceanic regions \cite{haza2018drogue}. These floating sensors, which are designed to collect surface current data, have been extensively studied, and the resulting data have been analyzed in numerous publications \cite{chaturvedi2020mathematical,gulakaram2018role}. Similarly, satellite-tracked drifters have been used in semi-enclosed seas, like the Adriatic Sea, to examine circulation patterns \cite{poulain2001adriatic, ursella2006surface}. Equipped with Global Positioning System (GPS), General Packet Radio Service (GPRS), and Very High Frequency (VHF) signals, these drifters offer valuable insights into ocean circulation. Unfortunately, they can drift outside the target areas, resulting in scattered measurements across vast regions. Additionally, even thouth drifter data can be useful, it does not provide a complete picture of the circulation across the entire domain, as it only offers point-based measurements.

In order to enhance model accuracy and incorporate Lagrangian data more effectively, two primary methods have been explored for reconstructing and integrating Lagrangian data into models. The first method estimates velocities by calculating the ratio of observed position changes over time, and then uses these velocities to adjust model predictions \cite{hernandez1995mapping}. The second approach employs an observational operator based on the particle advection equation, optimizing the Eulerian velocity field by minimizing discrepancies between observed trajectories and model results \cite{molcard2003assimilation}. Earlier studies \cite{bennett1987accuracy} indicated that Euler methods could introduce trajectory errors in non-uniform flow fields. Therefore, modern pollutant prediction models rely on more physically informed formulations. This approach maintains a consistent Eulerian framework, calculating slick thickness using layer-averaged Navier–Stokes equations and simulating pollution dynamics with the advection–diffusion equation \cite{tkalich2006cfd}.

A key challenge with Lagrangian drifter data is that the drifters move with ocean currents, leading to an uneven distribution, with drifters often clustering or drifting out of areas of interest. While Lagrangian data provides trajectories over space and time, research often shifts from trajectory analysis to reconstructing Eulerian velocity fields, with many studies utilizing drifter data for this purpose \cite{rao1981method,eremeev1992reconstruction,cho1998objectively}. While satellite measurements are valuable, they face challenges such as limited availability and occasional atmospheric interference. As a result, high-frequency (HF) radar has been used for near real-time surface velocity measurements, often serving to validate ocean current models \cite{solano2018development, marmain2014assimilation}. Still, HF radars has its limitations, such as shallow depth penetration, vulnerability to interference, difficulties in achieving high spatial resolution, and the need for substantial financial investment.

To overcome most of these challenges, machine learning (ML) has emerged as promising alternative to traditional interpolation and optimization methods for flow field predictions \cite{ghalambaz2024forty}. Studies such as \cite{grossi2020predicting, song2024developing} have utilized artificial neural networks (ANNs) to analyze temporal patterns in drifter trajectories and predict long-term movement, reducing errors in drifter models. Additionally, deep learning (DL) has been applied to surrogate modeling of fluid flows \cite{sun2020surrogate,tang2021deep}, enabling fast and efficient estimations without the reliance on extensive computational fluid dynamics (CFD) simulations.

Given the complexity of surface flow reconstruction and pollution spread modeling, approximating sea surface velocity could provide a significant advantage for rapid response. Therefore in this work, we propose a methodology for estimating the sea surface velocity field across an entire domain using only sparse and scattered drifter data, while maintaining the spatial complexity of surface flow. A fusion of two simplified two-dimensional models, combining bounded and open domains, is used as a surrogate to efficiently replicate submesoscale patterns. The bounded domain represents the realistic region of interest, incorporating all relevant elements such as the coastline and inflows/outflows. The open domain, defined as a fully open circular zone, captures the influence of external factors on the flow. Both surrogates are solved separately and then combined, allowing the simplified 2D flow model to account for broader flow dynamics. This approach enhances computational efficiency by reducing the need for extensive measurements and shortening simulation times. To prioritize speed, the initial flow model excludes factors such as wind, waves, tides, and temperature variations, which are later incorporated into a secondary surrogate. Through a hybridization process, the secondary surrogate compensates for the simplifications made in the primary model.

The paper is organized as follows: Section 1 introduces the motivation for tracking and predicting passive scalar fields, along with a review of related work. Section 2 explains the two-dimensional fusion flow model. Section 3 presents the formulation of the model fitting problem. In Section 4, the optimization procedure is described, including initialization and the implementation of an adaptive diffusion coefficient. Section 5 covers the acquisition of measurements for flow reconstruction and the prediction of passive scalar field movement. Section 6 assesses the accuracy of passive scalar field advection. Section 7 discusses the results from synthetic cases and the provided experiment. Section 8 highlights the limitations and provides a discussion. Finally, Section 9 concludes the paper.

\section{Simplified two-dimensional flow model}

Simulating transient ocean behavior requires substantial computational resources and extended simulation times. To address this, we adopt a fused flow model based on steady-state, incompressible flow. This approach deliberately omits dynamic factors like wind, waves, tidal fluctuations, and temperature changes, focusing on a fast yet sufficiently accurate approximation of sea surface flow. However, these environmental influences are accounted for through a fusion approach.

This simplified methodology offers an efficient alternative to traditional techniques. It eliminates the need for extensive datasets, complex interpolations, and domain models, thereby significantly reducing computational costs and processing time. By leveraging CFD simulations, the model captures essential fluid flow physics, enabling the calculation of accurate velocity fields and supporting advection/diffusion processes for passive scalar distribution.

The proposed approach is particularly useful in scenarios that require rapid computation of passive scalar field distributions, such as pollutant dispersion or object tracking. The model’s design leverages the quasi-transient nature of oceanic flows—where changes are gradual rather than abrupt—making it an effective tool for real-time or near-real-time applications. It strikes a balance between computational efficiency and the ability to capture transient-like flow dynamics, offering a practical solution for rapid response scenarios without the complexity of fully transient simulations.

\subsection{Steady-state 2D flow model}
\label{subsec:Steady-state_2D_flow_model}

The proposed steady-state flow model governs the movement of fluids at low to medium velocities within a connected computational domain $\Omega \subset \mathbb{R}^2$ , utilizing the incompressible Navier-Stokes equations \cite{gunzburger2012finite, lions1996mathematical, kronbichler2009numerical}:
\begin{equation}
	\rho\left(\mathbf{u}\cdot\nabla\right)\mathbf{u}-\nabla \cdot \left(2\mu\epsilon(\mathbf{u})\right)+\nabla p = \rho\mathbf{f}
	\label{eq:ns_equation}
\end{equation}
\begin{equation}
	\nabla \cdot \mathbf{u} = 0
	\label{eq:compressibility_equation}
\end{equation}

The vector $\mathbf{u}$ represents the velocity of the fluid, while $p$ denotes its dynamic pressure. The rank-2 tensor $\epsilon(\mathbf{u}) = \frac{1}{2} \left( \nabla \mathbf{u} + (\nabla \mathbf{u})^\top \right)$ defines the viscous stress tensor in an incompressible Newtonian fluid. In this context, $\rho$ stands for the fluid density, and $\mu$ is the fluid's dynamic viscosity.  The assumption of incompressibility implies that density remains constant. The term $\mathbf{f}$ accounts for external forces acting on the fluid.

To accurately represent the interactions between the observed section of the sea and the surrounding marine environment, a a specific set of boundary conditions is applied. At the boundary, tangential velocities and pressure values are prescribed. The proposed combination must not be randomly assigned; their range should align with realistic conditions to replicate the physical characteristics of actual flow.

In submesoscale oceanic regions such as the Adriatic Sea, surface currents exhibit significant variability due to interactions between mesoscale and smaller-scale processes. To capture these dynamics, numerical simulations require appropriate boundary conditions that account for realistic velocity magnitudes. Surface velocities in the Adriatic Sea, derived from radar data, infrared satellite imagery, and numerical simulations, range from below below 0.1 m/s to above 0.5 m/s \cite{cosoli2013surface,notarstefano2008estimation, bolanos2014modelling}. The average surface velocities in most areas are typically between 0.1 and 0.2 m/s. Given this variability, the boundary conditions for pressure and tangential velocity are set to ensure a velocity magnitude of up to 0.5 m/s within the domain. With the tangential velocity specified, and the internal velocity magnitude constraint, the normal velocity component can be derived and used to calculate the total pressure at the boundary: 

\begin{equation}
	p = p_0 - 0.5 \rho u^2
	\label{eq:pressure_calculation}
\end{equation}

where $p$ denotes the static pressure, $p_0$ is the total pressure, $\rho$ represents the fluid density, and $u$ is the normal velocity. 

After setting the tangential velocities and total pressure values, the next step is to establish the velocity and pressure profile functions. This process involves interpolating the tangential velocity and pressure values along the boundary while ensuring that the values at the coastline edges are set to zero. The procedure is updated during the optimization process until the boundary condition generates a velocity field that aligns with the velocity measurements from the drifter.

\subsection{Turbulence}
Turbulence is modeled using the \emph{k-$\omega$ shear stress transport} model \cite{ferziger2012computational}, which combines the benefits of both k-omega and k-epsilon models to improve the precision and robustness in simulating complex turbulent flows. This hybrid approach splits the flow domain into two regions: the near-wall and outer regions. For the near-wall region, the model uses a wall-function method to accurately capture turbulence near the wall. In the outer region, the model behaves like a free-stream model, ensuring reliable turbulence predictions away from the wall \cite{menter1992improved}. Additionally, it features an improved near-wall treatment, as k-$\omega$ model effectively resolves the boundary layer region. The turbulence variables  ($k$, $\omega$) are determined using:
\begin{equation}
	k = \frac{3}{2}(\left|\mathbf{u}\right|I)^2
	\label{eq:turbulence_kinetic_energy}
\end{equation}
\begin{equation}
	\omega = \frac{k^{0.5}}{C_\mu^{0.25} L}
\end{equation}\\
where $k$ is turbulence kinetic energy,  $I$ is  turbulence intensity, $\omega$  is the specific dissipation rate, $C_\mu$ is a turbulence model constant and equals 0.09 while $L$ is the turbulent length scale.

\subsection{Fusion model approach}
\label{subsec:Fusion_model_approach} 

Simulating realistic movement of a passive scalar within a specific domain - where environmental factors play a significant role - is often computationally demanding and challenging to implement using conventional CFD approaches. To overcome these limitations, a fusion based approach that combines two two-dimensional flow surrogates is adopted. The resulting fusion model serves as a surrogate for submesoscale dynamics and incorporates a correction offset to partially account for flow simplifications.

\begin{figure}[htbp]
	\centering{\includegraphics[width=\linewidth]{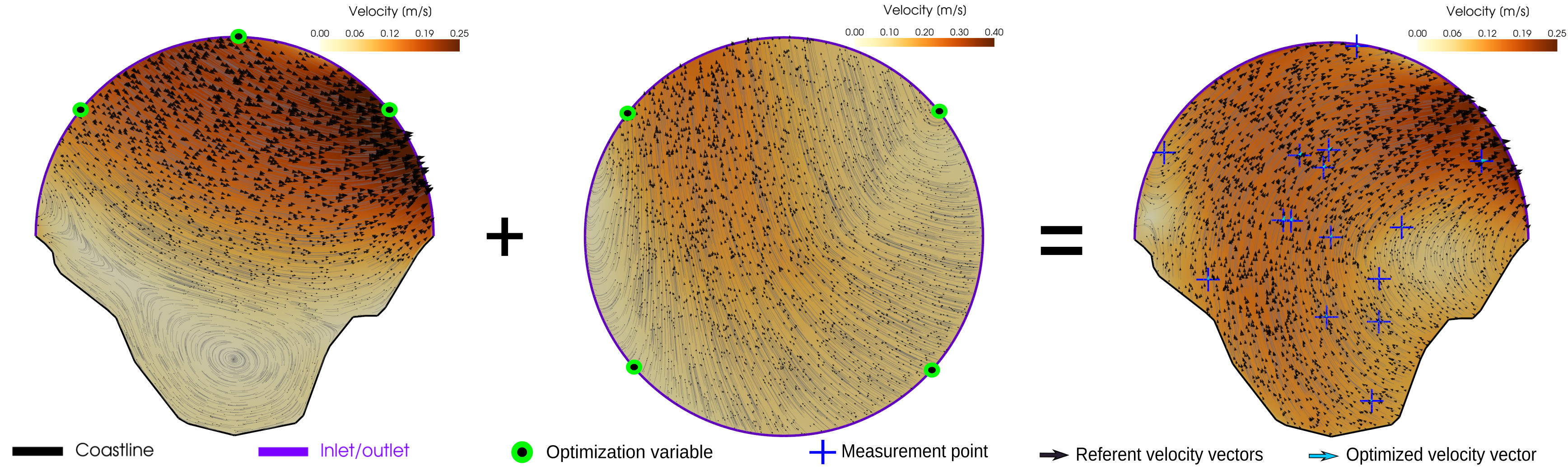}}
	\caption{The figure presents the methodology for running two simultaneous CFD simulations, with the bounded domain as the initial computational domain inside a larger open domain without a coastline, where this coastline-free domain forms the basis for non-uniform open domain flow. By combining the velocity fields from both simulations, the resulting flow provides a more realistic representation of surface dynamics.}
	\label{fig:combined_mesh}
\end{figure}

As noted, the fusion approach combines two distinct domains. The bounded domain is defined for the realistic region of interest and incorporates all relevant domain elements i.e. coastline and inflows/outflows. The open domain, which encompasses the bounded domain, is defined for a fully open circular zone in which external factors act. Both domain flows are solved separately and subsequently combined, allowing the simplified 2D flow surrogate to partially account for broader flow dynamics (Figure~\ref{fig:combined_mesh}).

The open domain uses four control points (eight additional variables for the optimization vector) to capture environmental variations, which are incorporated into the flow approximation to enhance the overall accuracy for the entire domain. By integrating velocity fields from both the bounded domain and open domain simulations, the open domain adjusts the flow field to better reflect real-world conditions. To achieve this superposition, velocity values at the node locations within the bounded domain are first extracted. These corresponding velocity values are then retrieved at the same locations in the open domain, ensuring alignment between the two fields. Finally, the velocity fields are summed, effectively incorporating corrective adjustments into the overall flow representation
\begin{equation}
	\mathbf{u}_{tot} = \mathbf{u}_{bounded} + \mathbf{u}_{open}
	\label{eq:flow_fusion}
\end{equation}
where $\mathbf{u}_{bounded}$ and $\mathbf{u}_{open}$ are velocity fields obtained for bounded and open flow models, respectively.

To incorporate the non-uniform open domain flow into the bounded domain, surface velocities from the open domain simulation are interpolated onto the mesh points of the bounded domain. This ensures that variations in wind speed and direction, as influenced by the surrounding environment, are accounted for in the flow reconstruction. By summing the velocity fields from both simulations, a more dynamic and accurate flow model is obtained, reflecting the complex environmental interactions within the domain. This fusion approach enables adaptive, real-time adjustments to the flow field, improving the reconstruction of the passive scalar field. By incorporating these environmental effects, the simulation is better equipped to handle dynamic conditions, more closely mirroring real-world scenarios.

\subsection{Passive scalar transport}
\label{subsec:Passive_scalar_transport}  

Along with the flow modeling, we incorporated a scalarTransport object into the simulation to govern the advection and diffusion of the passive scalar field at each time step. The passive scalar field is advected on the velocity field of fused flow \eqref{eq:flow_fusion} as described by the following scalar transport equation:
\begin{equation}
	\frac{\partial s}{\partial t} + \mathbf{u}_{tot}\nabla s - D\nabla^2s=0
	\label{eq:scalar_transport}
\end{equation}
where $s$ is the scalar field, $\mathbf{u}$ is the velocity field, $t$ represents time and $D$ s the diffusion coefficient. The diffusion coefficient is determined using:
\begin{equation}
	D=\alpha_1v + \alpha_2v_t
	\label{eq:diffusion_coeff}
\end{equation}
where $\alpha_1$ and $\alpha_2$ are constants and are equal to $1e^{-9}$ and 0.714 respectively. $v_t$ represents turbulent viscosity. Kinematic viscosity $v$ is set to $1e^{-6}$ $m^2/s$. 

This scalar transport equation is crucial for replicating the realistic movement of the passive scalar field throughout the domain, ensuring that the behavior of the scalar field is appropriately modeled in relation to the evolving flow dynamics. This combined approach allows for the accurate representation of both the fluid flow and the transport of the passive scalar field, ensuring the robustness of the transient flow simulation.

\subsection{Numerical implementation}
\label{subsec:Numerical implementation}

The computational domains in the considered test cases are designed to be either synthetic, for controlled experiments, or to represent a realistic geographical region for practical applications. For realistic regions, data acquisition is performed using Sentinel Hub’s polygon extraction tool, accessible via \href{https://www.sentinel-hub.com/}{Sentinel Hub}. The Sentinel Hub L2A NDWI (Normalized Difference Water Index) visualization is utilized, focusing on NDWI TIFF images of the specified region. These images enable precise identification of water bodies and coastlines.

To define the domain boundaries, polygons corresponding to the region of interest are created. The process involves extracting coastline points based on NDWI indices, which are then refined and smoothed to ensure accuracy. Following the polygon extraction, a stereolithographic (STL) model is
created, which serves as a geoemtric baseline to create a computational mesh.

Subsequently, a two-dimensional numerical mesh is generated using the cfMesh library \cite{juretic2015cfmesh}. This mesh is utilised to define a numerical model/simulation/problem using open-source CFD software OpenFOAM \cite{openfoam} for conducting simulations. This workflow ensures precise representation of geographical features and enables the simulation of realistic fluid dynamics within the specified domain.

Velocity at the coastline is defined as a no-slip (Dirichlet) boundary condition. When the fluid exits the domain at a boundary face, a Neumann boundary condition is applied to the velocity, i.e. the fluid velocity at the boundary is extrapolated from the velocity inside the domain. For the fluid entering the domain, the open sea boundary is set to a Dirichlet condition, with the velocity computed based on the flux in the patch-normal direction. Additionally, we specify tangential velocities, as the incoming flow may not be perfectly aligned with the inlet boundaries. Defining the tangential velocity allows for more realistic inlet flow conditions, accounting for potential swirl or tangential motion of the fluid. 

The pressure at the open sea boundary is modelled using a Dirichlet Dirichlet boundary condition with a specified range of values, while at the coastline, a Neumann boundary condition is applied. To maintain a consistent pressure field within the domain, a reference cell is selected and assigned a pressure value of zero. This reference point is then used to compute the pressure gradient across the entire domain.

Initial values for turbulent kinetic energy (k) and specific rate of dissipation ($\omega$) are estimated according to the defined equations. For walls, appropriate wall functions are used. A summary of the boundary conditions for all test cases is provided in Table \ref{tab:boundary_conditions}.

\begin{table*}[!htb]
	\footnotesize
	\centering
	\caption{An overview of the boundary conditions employed}
	\label{tab:boundary_conditions}
	\begin{tabularx}{\linewidth}{Xrrl} 
		\\
		
		\textbf{Field}							& Inlet/Outlet	& Coastline				\\
		\hline 	
		
		$u$										& pressureInletOutletVelocity	& noSlip    					\\
		
		$p$										& totalPressure			& zeroGradient						\\
		$k$										& fixedValue 					& kqRWallFunction	\\
		$\omega$								& fixedValue 			& omegaWallFunction			\\
		
		\\

	\end{tabularx}

\end{table*}

To replicate transient flow in this study, we opted to use the simpleFoam steady-flow solver within OpenFOAM, performing multiple short steady-flow simulations to approximate transient behavior. This solver utilizes the semi-implicit method for pressure-linked equations (SIMPLE)  \cite{patankar1983calculation}. For turbulence modeling, the k$\omega$-SST model is employed to accurately capture the flow characteristics.

In the OpenFOAM simulations conducted for steady-state flows, second-order accuracy was predominantly utilized, employing second-order gradient and Laplacian schemes. To enhance stability in areas with steep gradients, first-order schemes (Gauss upwind) were applied specifically to divergence terms associated with convective transport. Time derivatives were resolved using default second-order schemes, and linear schemes were adopted for interpolation. The meshWave method was used to compute distances to the nearest wall. The boundary conditions and simulation setup remained consistent across all test cases.

Further details on the numerical implementation for both bounded and open domain cases, including information on the numerical grid, cell distribution, discretization, and modeling schemes, are available in The Open Science Framework repository: \url{https://osf.io/wjsb2/}

\section{Model fitting problem formulation}

The proposed methodology employs an optimization algorithm that iteratively refines the values in the optimization vector to reduce the error, i.e., to minimize the discrepancy between the simulated flow field and the reference flow. Consequently, the accuracy and reliability of the flow field reconstruction are highly dependent on the optimization vector $\mathbf{b}$, which includes the tangential velocity and pressure values at the boundary control points:
\begin{equation}
	\mathbf{b} = \left(\mathbf{u}_{t,1}, p_1, \ldots,\mathbf{u}_{t,n_{CP}}, p_{n_{CP}} \right)^T
\end{equation}
where $n_{CP}$ represents the number of boundary control points. 

To capture realistic surface flows and address the high variability of surface currents, particularly in submesoscale domains, the optimization variable bounds are set to -0.5 to 0.5 $m/s$ for tangential velocity and -0.05 to 0.05 $m^2/s^2$ for pressure at the boundary control points, as disucssed in \ref{subsec:Steady-state_2D_flow_model}. These bounds are defined for numerical stability, with initial optimization candidates randomly distributed within this range, as they do not significantly affect the qualitative outcomes. It is worth noting that the final values for total pressure and tangential velocity at the boundary may deviate slightly from the assigned values due to adjustments based on the results of the Navier-Stokes equations within the internal domain. After defining boundary control point values, cubic spline interpolation is used to compute velocity and pressure values for each cell along the boundary. During optimization, the velocity profile is iteratively updated as the optimization vector evolves until the output error aligns with the target fitness.

\subsection{Objectives}\label{objectives}
The measurements are defined as data points, each caracterized by its coordinate and associate velocity vector. For each evaluation, a complete OpenFOAM case is created to compute the velocity field across the entire domain. The velocity vectors at the measurement point coordinates, representing drifter positions, are then extracted and treated as reference values to match at those specific locations. To quantify the deviation, the cost function calculates the drifter error, $\epsilon_d$, which aggregates the discrepancies at these discrete points (drifters) in meters per second, as defined by the following equation:

\begin{equation}
	\epsilon_d(\mathbf{b})  = \dfrac{1}{n_{MP}} \sum_{i=1}^{n_{MP}} (\mathbf{u}_{r_i} - \mathbf{u}_{s_i}(\mathbf{b}))^2 
	\label{eq:drifter_error}
\end{equation}
where  $n_{MP}$ represents the number of measurement points, $\mathbf{u}_{r_i}$ denotes the referent velocity vector,  while $\mathbf{u}_{s_i}$ is the simulation velocity vector at the measurement point location $\mathbf{s}_i$

\subsection{Constraints}\label{constraints}
To obtain feasible solutions within this simulation-based optimization framework, it is crucial to establish suitable constraints. Constraints are tied to the simulation residuals, which guide the optimization process towards achieving the desired fitness and ensuring numerically stable results. When the residuals are below the specified thresholds, the constraints are deemed satisfied. If the residuals exceed the limits, a penalty is imposed for failing to meet the criteria.

The pressure residual constraint, given by:
\begin{equation}
	r_p(\mathbf{b}) \leq 1e^{-3}
	\label{eq:p_constraint}
\end{equation} helps ensure consistent pressure values during the optimization process, preventing pressure imbalances that could result in unrealistic behavior. The velocity residuals ensure controlled and physically realistic fluid motion and are defined for both velocity components:
\begin{equation}
	r_{u_{x}}(\mathbf{b}) \leq 1e^{-4}
	\label{eq:vel_x_constraint}
\end{equation} 
\begin{equation}
	r_{u_{y}}(\mathbf{b}) \leq 1e^{-4}
	\label{eq:vel_y_constraint}
\end{equation}

The \emph{k} residual constraint, defined as:
\begin{equation}
	r_k(\mathbf{b}) \leq 1e^{-4}
	\label{eq:k_constraint}
\end{equation} ensures that the turbulent kinetic energy remains within acceptable bounds. The $\omega$ residual constraint, set as:

\begin{equation}
	r_\omega(\mathbf{b}) \leq 1e^{-4}
	\label{eq:omega_constraint}
\end{equation} limits the specific dissipation rate of turbulence, maintaining alignment with the fundamental physics. These constraints collectively guide the optimization process, ensuring physical realism, stability, and relevant fluid dynamics representation. All five constraints are assessed across all test cases.

\section{Optimization procedure}

The selection of the most suitable optimization algorithm for similar flow field reconstruction method was previously addressed in \cite{jakac2024approximation}, where the performance, efficiency, robustness, and scalability of various methods were evaluated, leading to the adoption of PSO as the preferred approach for this type of modeling. Although the authors noted that optimization outcomes are case-dependent, they observed similar results across different test cases due to the synthetic nature of the simulations, despite variations in data sources.

As outlined in \cite{jakac2024approximation}, the mean square difference of the velocities at measurement points, denoted as $\epsilon_d$, is used as the fitness function in all optimization tests. The optimization is thus defined as follows:

\begin{equation}
	\begin{aligned}
		& \underset{\mathbf{b}}{\text{minimize}}
		& & \epsilon_d(\mathbf{b}) = \dfrac{1}{n_{MP}} \sum_{i=1}^{n_{MP}} (\mathbf{u}_{r_i} - \mathbf{u}_{s_i}(\mathbf{b}))^2\\
		& \text{subject to}
		& & \mathbf{b}_l \mathbf{\leq b} \leq \mathbf{b}_u
	\end{aligned}
\end{equation}

Convergence of the optimization process is considered to be achieved when the drifter error threshold, $\epsilon_d$ = 1$e^{-4}$, is reached, corresponding to a drifter velocity error in $m/s$. All reconstruction results for which these threshold is met are deemed satisfactory.

\subsection{Initial flow fitting}
The idea behind recreating passive scalar field advection and diffusion based on a stationary reconstructed flow is to advect and diffuse the passive scalar field according to the times at which the measurements are collected. As the movement of the passive scalar field is heavily influenced by the initial reconstructed flow, and since errors accumulate throughout the process, it is crucial to extend the optimization duration for the first flow reconstruction. To minimize this error accumulation, the initial flow reconstruction is given a duration five times the time step at which the measurements are obtained. This ensures the flow is accurately reconstructed before being used in subsequent steps. After the first reconstruction, the optimization time is shortened to match the time step of new measurements, which are used to update the flow for further advection and diffusion calculations. Since the flow field is not expected to change significantly between time steps, this approach—starting with a longer reconstruction and followed by shorter ones based on updated measurements ensures accurate reconstruction of the transient flow dynamics. As a result, the movement of the passive scalar field is reliably replicated by using consequtive stationary flow reconstructions

\subsection{Optimization initialization}\label{opt_initialization}

The optimization process seeks to find the best solution by iteratively adjusting the optimization vector. The optimization variable bounds are defined as -0.5 to 0.5 $m/s$ for tangential velocity and -0.05 to 0.05 $m^2/s^2$ for pressure at the boundary control point, as detailed in subsection \ref{subsec:Steady-state_2D_flow_model}. This means that each simulation starts with internal field values initialized to zero, while the boundary conditions differ. However, certain configurations of optimization vector values may sometimes result in simulations that struggle to converge or take considerably longer to reach convergence, increasing the overall time required for optimization.

Given that the flow field is not expected to change significantly between time steps, the initial bounds for optimization variables are applied only during the first flow reconstruction. For subsequent reconstructions, the optimisation bounds are updated and limit the search range within 30\% of the current best optimisation result i.e:

\begin{equation}
	\begin{aligned}
		&  \mathbf{b}_l^* = \mathbf{b}_{opt} - 0.3(\mathbf{b}_u - \mathbf{b}_l)\\
		& \mathbf{b}_u^* = \mathbf{b}_{opt} + 0.3(\mathbf{b}_u - \mathbf{b}_l)\\
	\end{aligned}
\end{equation}

where, $\mathbf{b}_l^*$ and $\mathbf{b}_u^*$ represent the updated optimization bounds for the optimization variables, while $optX$ refers to the optimization variables from the best solution of the previous optimization. This refinement reduces the search space for subsequent iterations, enhancing the optimization process by improving optimization/fitting convergence and concentrating on the most favorable parameter ranges for each time step.

Additionally, to further enhance optimization efficiency and decrease simulation time, a flow-based initialization method, inspired by \cite{jakac2024approximation}, is applied. In this approach, the optimization process starts with the best solution from the previous step, using its internal field values and optimization variables to initialize the new simulation. By combining different initialization approaches and the reduced search space concept, the overal number of iterations is reduced, leading to quicker simulations and overal faster convergence.

\subsection{Adaptive diffusion coefficient based on reconstruction flow error}\label{adaptive_diff}

To mitigate errors in flow reconstruction and their effects on the advection-diffusion of the passive scalar, we employ the mean square displacement (MSD) of a Brownian particle to compute a compensating diffusion coefficient. The MSD quantifies the average displacement of a particle undergoing diffusion over time and is expressed as:

\begin{equation*}
	E^2(t) = 2 \cdot n_{dim} \cdot D \cdot t,
\end{equation*}

where \(E\) is the mean displacement, \(n_{dim}\) is the number of spatial dimensions (\(n_{dim} = 2\) in this case), \(D\) is the diffusion coefficient, and \(t\) is the time interval.

Reconstruction errors introduce deviations between the reconstructed flow field and the referent flow field, resulting in a net displacement over a time step \(\Delta t\). This displacement is calculated using the reconstruction error \(\epsilon_d\) incorporated into the MSD relationship for two-dimensional diffusion to determine the compensating diffusion coefficient:

\begin{equation*}
	D_{c} = \frac{(\sqrt{\epsilon_d} \cdot \Delta t)^2}{4 \cdot \Delta t}.
\end{equation*}

Here, \(\sqrt{\epsilon_d} \cdot \Delta t\) represents the mean displacement associated with the drifter error over the time step \(\Delta t\). This compensating diffusion coefficient quantifies the uncertainty introduced by inaccuracies in flow reconstruction.

Therefore, the adaptive diffusion coefficient for the reconstructed flow, \(D_{adp}\), can be expressed as:

\begin{equation*}
	D_{adp} = D_{base} + D_{c}.
\end{equation*}

In this formulation, \(D_{base}\) denotes the diffusion coefficient of the referent flow, while the adaptive diffusion coefficient \(D_{adp}\) is recalculated at each time step using newly acquired measurements. This adjustment compensates for reconstruction errors, ensuring that the advection-diffusion process more closely aligns with the behavior of the passive scalar field under the referent flow conditions.

\section{Acquiring measurements}
\subsection{Measurement point advection}\label{sp_advection}
As drifters move through the domain under the influence of currents, we opted to model their movement as a quasi-transient advection problem. This process involves advecting the positions of the measurement points, or drifters, within the domain. The positions are updated based on the reconstructed velocity field, with the movement of each drifter determined by the velocity field from the preceding simulation step. The position is then adjusted using the advection equation:

\begin{equation}
	\mathbf{x}\left(t+\Delta t\right) = \mathbf{x}(t) + \mathbf{u}(\mathbf{x}(t))\Delta t
	\label{eq:location_advection}
\end{equation}

where $\mathbf{x}(t)$ represents the position of the measurement point at time $t$, $\mathbf{u}(\mathbf{x}(t))$ is the velocity at that position, and $\Delta t$ is the time step over which the advection occurs. This approach ensures that the measurement points are corrected according to the flow dynamics at each time step, enabling the tracking of their positions over time. The advection of the measurement points is critical for obtaining measurements at new locations, thereby allowing for the reconstruction of the velocity field and capturing the evolution of the flow. By iterating this process across different flow states, the movement of the drifters is effectively simulated, providing a reliable representation of transient flow dynamics in the domain.

\subsection{Absent measurements}\label{absent_measurements}

In real-world scenarios, absent measurements can arise for several reasons. One common cause is that drifters may drift outside the monitored area, leading to a loss of data. Additionally, signal interruptions can occur, particularly when drifters travel over large distances or encounter environmental factors such as rough sea conditions or physical obstructions that disrupt communication, compounded by the limitations of the tracking technology. Despite these potential data deficiencies, the proposed flow reconstruction method is designed to function independently of the number of available measurements at any given time. By leveraging the inheritance approach discussed in earlier sections, the method compensates for missing data, as the flow field is not expected to change drastically between time steps. This approach allows for effective reconstruction of the flow field, even in the presence of scattered or incomplete measurements, ensuring that the overall accuracy of the flow approximation remains reliable.

\section{Assessment of passive scalar field advection accuracy}

To evaluate the accuracy of passive scalar advection, we compare two advection/diffusion processes: one driven by referent flow which represents the referent scalar field movement we want to replicated and the other simulated with quasi-steady flow assumptions. The comparison is based on the areas of advected/diffused probability by comparing the resulting scalar fields. Both scalar fields, which represent the passive scalar field, are initialized at the same location with identical values to ensure consistency.

To generate the referent advection process, we synthetically create referent flow cases with time-varying boundary conditions, solving the scalar transport equation alongside the referent flow to track passive scalar movement. These boundary conditions produce transient flow dynamics, which we aim to approximate using quasi-steady flow assumptions. Given the importance of accurate initial reconstruction, the first optimization step is extended to 5$t_{M}$ allowing sufficient time for convergence and minimizing errors that could propagate and compromise advection accuracy. Subsequent optimizations are performed within a $t_{M}$-long window, ensuring continuous updates to the flow field based on new measurements.

This iterative process enables the method to adapt to changing boundary conditions while approximating transient flows under quasi-steady assumptions. By reconstructing the velocity field at each $t_{M}$, the system effectively captures flow variations and provides a reliable basis for modeling passive scalar advection.

To assess the accuracy of passive scalar advection, we use two metrics. The first metric, called the intersection metric, quantifies the portion of the referent scalar field that is captured by the simulated scalar field. It is defined as:

\begin{equation}
	I = \frac{\int_{\Omega_{intrsc}} \phi_{sim} , d\Omega}{\int_{\Omega_{ref}} \phi_{ref} , d\Omega}
\end{equation}

where $\Omega_{intrsc}$ represents the intersection area between the simulated and referent passive scalar fields, and $\phi$ denotes the passive scalar values. The intersection metric is scaled between 0 and 1, where a value of 1 indicates perfect alignment between the simulated and referent scalar fields, meaning the reconstructed flow completely replicates the transient dynamics. A value closer to 0 indicates a greater discrepancy between the fields, highlighting the limitations of the reconstructed flow in capturing transient advection effects.

The second metric,called the coverage metric, measures the extent to which the simulated scalar field is covered by the referent scalar field. It evaluates how much of the referent probability field is represented by the simulated probability field. The coverage metric is calculated as:

\begin{equation} 
	C = \frac{\int_{\Omega_{intrsc}} \phi_{sim} \, d\Omega}{\int_{\Omega_{sim}} \phi_{sim} \, d\Omega} 
\end{equation}

A value of 1 indicates that the referent passive scalar field fully covers the simulated probability field, while a value closer to 0 indicates less coverage.

Both the intersection and coverage metrics provide a quantitative evaluation of how well the reconstructed passive scalar field (based on stationary flow) corresponds to the reference passive scalar field (based on transient flow). These metrics provide insight into the effectiveness of the flow reconstruction and the precision of the advection process.

\section{Results}

In this study, we aim to reconstruct transient flow with changing boundary conditions across multiple $t_{M}$ steps to better account for dynamic environmental variations and demonstrate the adaptability of the proposed method to evolving conditions. At each $t_{M}$  step, we perform steady-state flow reconstruction using measurements obtained from the referent flow, ensuring that the optimization continuously adapts the simulation for gradually evolving flows. Such approach addresses time complexity, while the usage of a fusion model helps mitigate spatial complexity. This iterative reconstruction process allows the simulated passive scalar field to more accurately reflect real-world transport dynamics, particularly in scenarios where transient flow characteristics significantly influence advection patterns.

To validate our methodology, we designed three distinct test cases, each within a different domain: a synthetic domain, a realistic domain, and a domain where we conducted experimental field measurements. These cases were selected to systematically evaluate our approach under varying conditions, ranging from idealized to real-world scenarios. The specific characteristics of each test case are summarized in Table \ref{tab:case_characteristics}. This diverse set of cases enables a comprehensive assessment of our method's performance across different environments and scales.

\begin{table*}[!htb]
	\footnotesize
	\centering
	\caption{Characteristics of the validation test cases}
	\label{tab:case_characteristics}
	\begin{tabularx}{\linewidth}{Xrrrl} 
		\\
		
		\textbf{Case}							 		    	& Simple bay	& Lošinj case	& Cres case					\\
		\hline 	
		
		Test case type 												    		& Synthetic 		& Realistic  	& Realistic			\\
		
		Domain area [$km^2$]										 			& 24.6 		& 96.49			& 55.62 					\\
		Number of boundaries 							 						& 1 		& 4		     	& 1						\\
		Total boundary length [$km$]											& 9.4 	    & 9.63			& 7.22					\\
		Coastline length [$km$]								 					& 9.1		& 57.12			& 31.74					\\
		Number of boundary control points 										& 5		    & 6		    	& 3						\\
		
		Max velocity in the domain [$m/s$]									 	& 0.25		& 0.4	     	& 0.5 				\\

		Number of cells 										& 4625					& 7530			& 15833			\\
		Average cell size [$m$]										& 73.02				& 113.21		& 59.28			\\
		\\

	\end{tabularx}

\end{table*}

\subsection{Simple bay}
This synthetic domain of 25 $km^2$ provides a controlled environment with predefined flow conditions, allowing for a precise assessment of the reconstruction method’s performance without of external uncertainties. The domain consists of a coastline and an inlet/outlet, which are designed to simulate realistic boundary interactions. 

\begin{figure*}[!h]
	\centering
	\includegraphics[width=\linewidth]{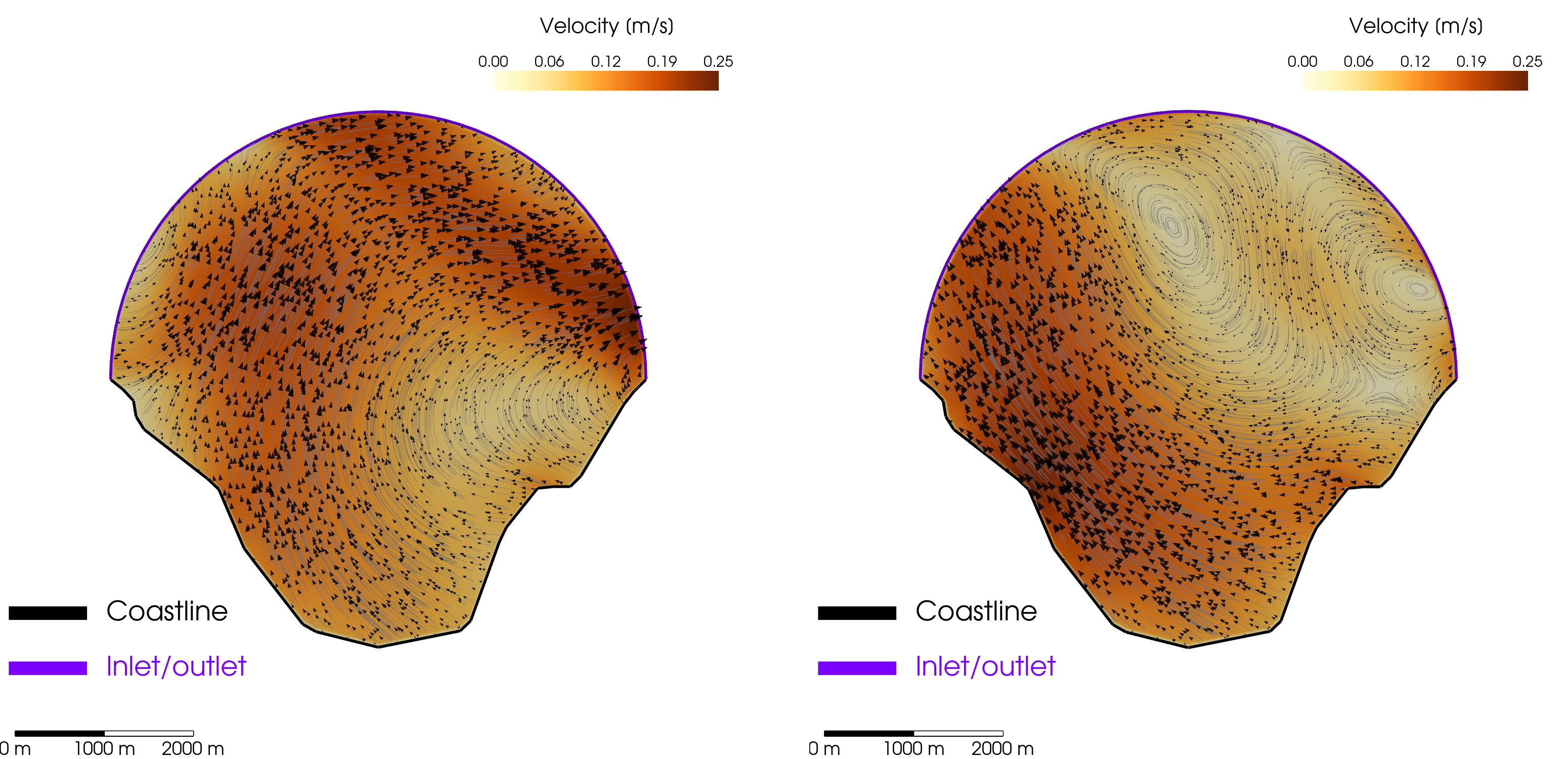}
	\caption{Figure represents one type of the evolving referent flow in the synthetic Simple bay case after 32400 seconds, highlighting the difference between the initial flow (left side) and the final flow (right side). As observed, the flow undergoes significant changes, demonstrating how the dynamics of the system can shift over time while making the flow reconstruction more challenging.}
	\label{fig:simple_bay_flows}
\end{figure*}

Fig. \ref{fig:simple_bay_flows} illustrates the evolution of the referent flow over 32400 seconds, highlighting variations in both direction and magnitude that we aim to reconstruct. By using this synthetic domain, we can isolate and assess the specific impact of boundary condition updates on the accuracy of passive scalar field reconstruction, focusing on the method's effectiveness in handling temporal flow changes. The case provides a clear validation of the methodology under controlled conditions, establishing a baseline for comparison with more complex real-world scenarios. While drastic flow changes would not typically occur over a short period in real-world cases, the synthetic nature of this domain allows us to test our methodology under extreme conditions and adaptability in reconstructing flow fields subjected to evolving boundary.

\begin{figure*}[!h]
	\centering
	\includegraphics[width=\linewidth]{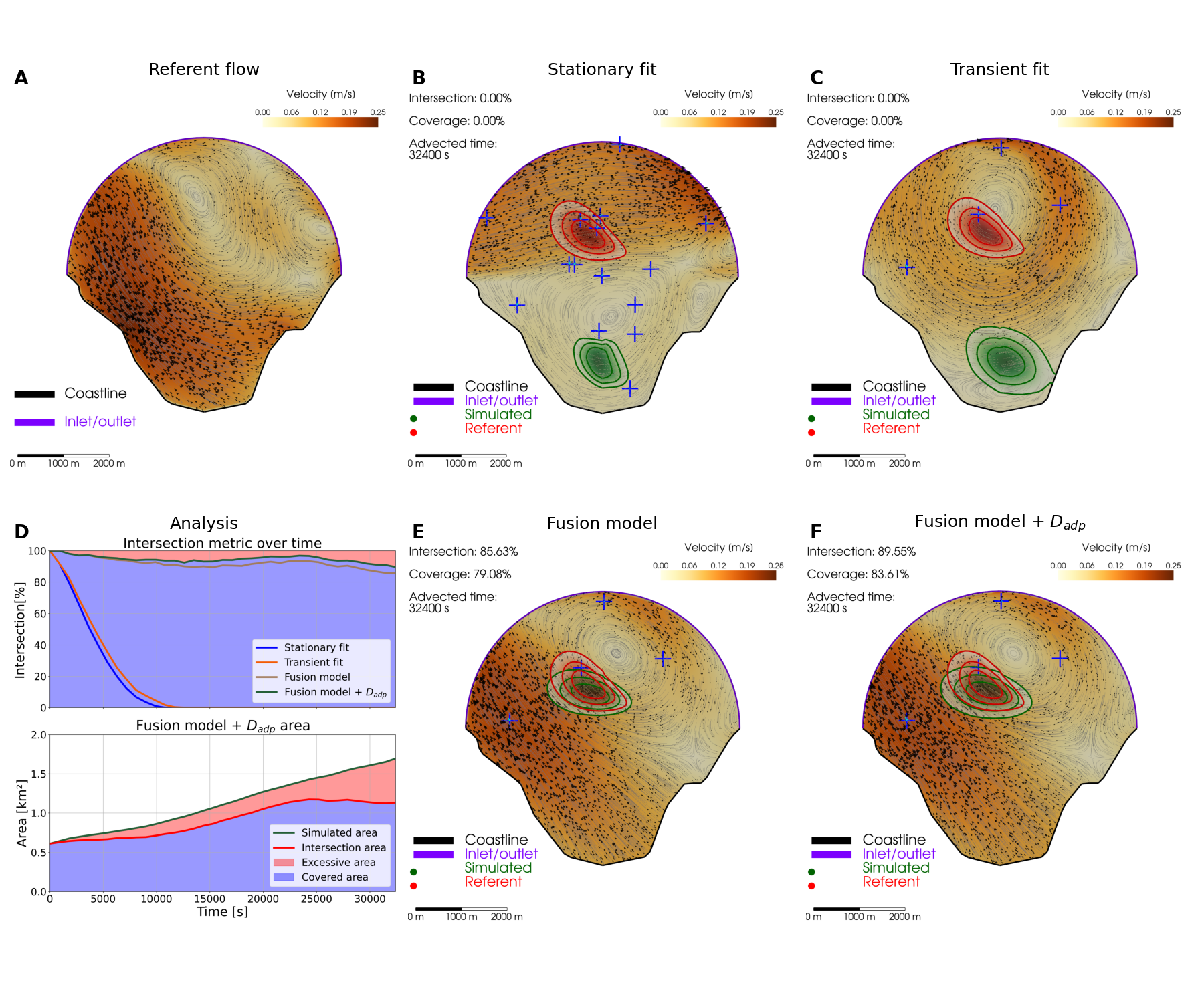}
	\caption{(\textbf{A}) evolution of referent flow after 32400 seconds, which we aim to reconstruct. (\textbf{B}) Stationary reconstructed flow based on 15 drifter measurements from the initial flow where the flow was reconstructed using initial state and conditions. (\textbf{C}) Flow reconstruction with periodic measurement corrections (updates) where fusion approach is not considered. (\textbf{D}) Intersection analysis comparing different approaches for passive scalar field advection simulation. (\textbf{E}) Flow reconstruction with updated measurements and fusion model. (\textbf{F}) Flow reconstruction with updated measurements, fusion model, and adaptive diffusion compensation to account for errors in flow reconstruction.}
	\label{fig:simple_bay_final_results}
\end{figure*}

Fig. \ref{fig:simple_bay_final_results} compares four different approaches for flow reconstruction and the resulting passive scalar advection. Due to the dynamic nature of the referent flow, the stationary-fit approach (\textbf{B}), which reconstructs the flow using only the initial 15 drifter measurements from the referent flow and then advects the passive scalar field over 32400 seconds without updates, led to significant errors, ultimately resulting in 0\% intersection with the referent field. The transient-fit approach (\textbf{C}), which reconstructs the referent flow at each $t_{M}$ measurement step across multiple steady-state flow optimizations, was also unable to capture the complex referent flow due to the limitations of the bounded CFD domain, again resulting in 0\% intersection. Notably, after 32400 seconds, only four drifters remained in the domain, meaning measurements were available from a limited number, further reducing the accuracy of the reconstructed flow.

To improve accuracy, the fusion model was introduced into the fitting process (\textbf{E}), enhancing the reconstructed flow accuracy and achieving an 86\% intersection. To further improve accuracy an adaptive diffusion coefficient (\textbf{F}) was introduced, with the fusion model producing even better results, reaching 90\% intersection after 32400 seconds.

The analysis of success (\textbf{D}) highlights the effectiveness of each approach in capturing the passive scalar field, providing insights into their respective strengths and limitations. Furthermore, the best-performing approach, which combines the fusion model and adaptive diffusion, not only maximized the intersection percentage but also optimized the balance between covered and excessive area, ensuring the most accurate passive scalar field reconstruction.

\subsection{Lošinj case}
In order to assess the proposed methodology for a realistic domain, we selected Unija Bay near the island of Lošinj, covering an area of 96.5 $km^2$. This domain features four distinct inlet/outlet areas, which can induce complex and dynamic flow patterns. Unlike synthetic, the realistic domain introduces naturally occurring flow variations influenced by external environmental factors, providing a more challenging test for the reconstruction methodology. 

\begin{figure*}[!h]
	\centering
	\includegraphics[width=\linewidth]{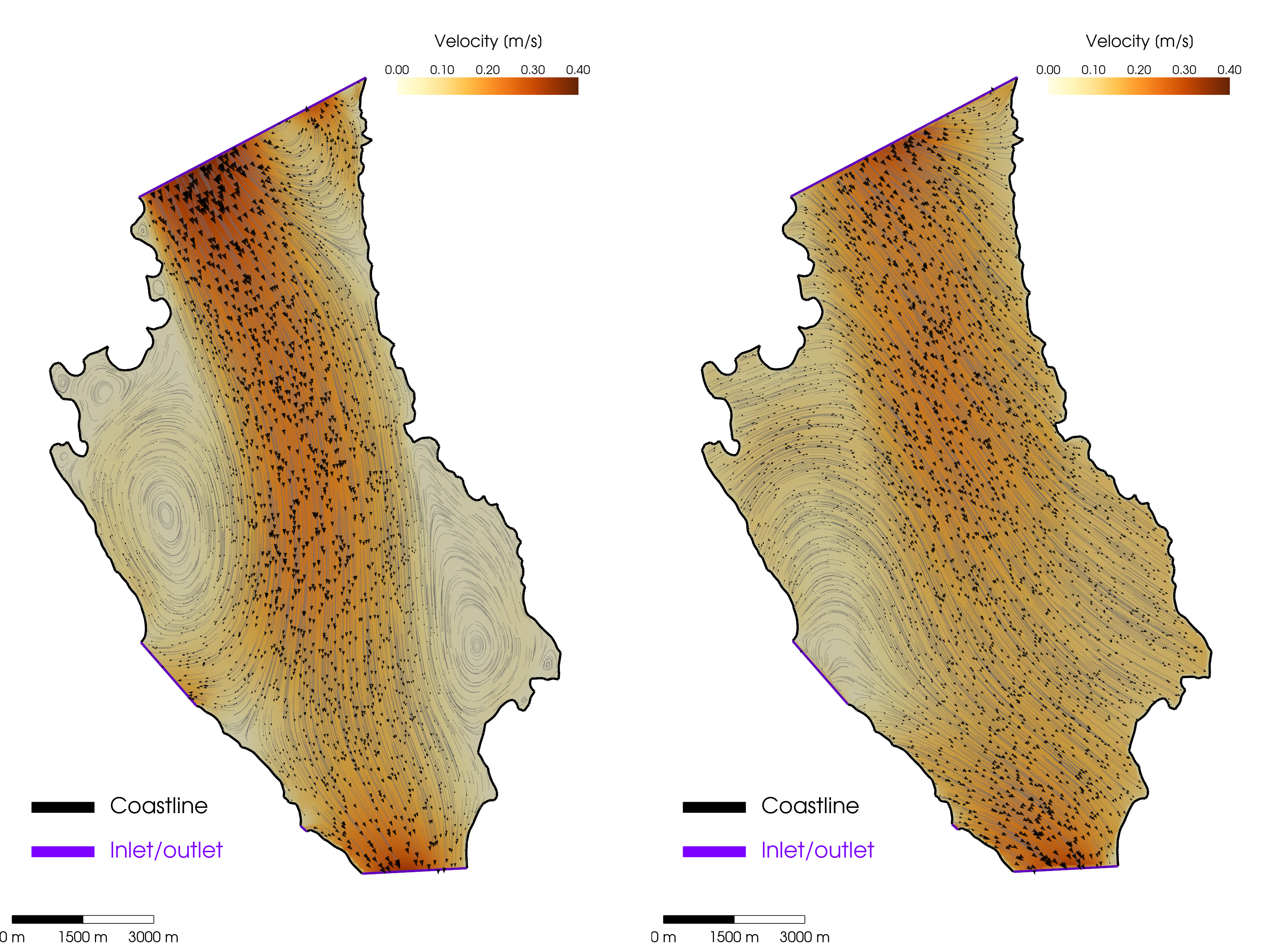}
	\caption{Evolution of the referent flow in the realistic Lošinj case after 36000 seconds, illustrating the differences between the initial flow (left) and the final flow (right). This case features four inlet/outlet areas, allowing for the formation of multiple vortices within the domain. Initially, the flow is directed toward the western and southern outlets, whereas after 36,000 seconds, it shifts toward the eastern coastline and the southern outlet.}
	\label{fig:losinj_flows}
\end{figure*}

To demonstrate the effectiveness of our methodology in a realistic domain, Fig.\ref{fig:simple_bay_final_results} compares four different approaches for flow reconstruction and the resulting passive scalar advection. As observed in the synthetic case, both the stationary fit (\textbf{B}) and transient fit (\textbf{C}) failed to replicate the referent passive scalar movement. Surprisingly, in this instance, the transient fit performed even worse than the stationary fit, indicating that optimization can sometimes converge to incorrect solutions.

Incorporating the fusion model (\textbf{E}) into the transient fitting significantly improved accuracy, achieving nearly 93\% intersection with the referent passive scalar field after 36,000 seconds. Further improvement was achieved by introducing an adaptive diffusion coefficient (\textbf{F}), which yielded an intersection metric of 94\% and a coverage metric of 100\%.

\begin{figure*}[!h]
	\centering
	\includegraphics[width=\linewidth]{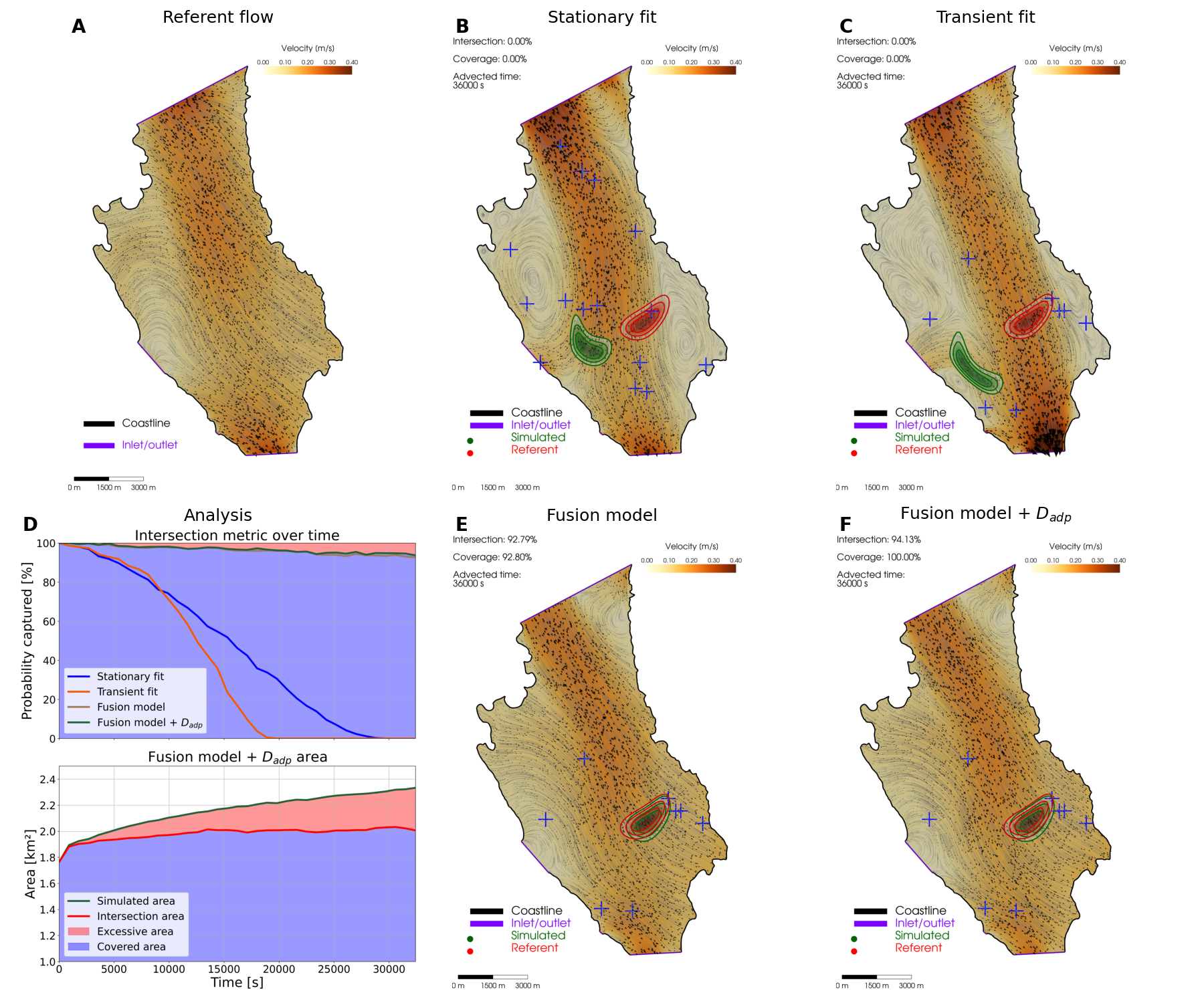}
	\caption{(\textbf{A}) The reference flow after 36000 seconds, predominantly directed toward the eastern coastline, a pattern commonly associated with the Jugo wind in this region. (\textbf{B}) Stationary flow reconstruction based on 15 drifter measurements from the initial flow, resulting in no intersection with the reference passive scalar field. (\textbf{C}) Transient flow reconstruction without the fusion model. An even greater discrepancy between the simulated and reference passive scalar field is observed. (\textbf{D}) Intersection and adaptive diffusion analysis, comparing different approaches for passive scalar advection. (\textbf{E}) Transient flow reconstruction incorporating the fusion model yields excellent results. (\textbf{F}) Flow reconstruction integrating fusion approach and adaptive diffusion compensation. Captured passive scalar field is now more in line with the referent state. 	}
	\label{fig:losinj_advection_results}
\end{figure*}

The success analysis (\textbf{D}) evaluates the effectiveness of each approach in capturing the passive scalar field, revealing a slight improvement with the inclusion of the adaptive diffusion coefficient and unexpectedly better performance of the stationary fit compared to the transient fit. Moreover, the best-performing approach—combining the fusion model with adaptive diffusion—covered excessive area. excessive area. However, the coverage metric indicates that this excess region contains a minimal trace of passive scalar, which can partially be attributed to numerial errors, and can thus be disregarded when predicting movement of the bulk of the passive scalar.

\subsection{Cres case}

To properly validate the predictive algorithm, a sea experiment was conducted in the larger area of Valun Bay on the island of Cres, located in the Kvarner Bay region of the northern Adriatic Sea, Croatia. This location covers an area of 5 km by 10 km in longitudinal and latitudinal directions and was selected for its balanced environmental conditions, making it well-suited to represent submesoscale flow. While more exposed than smaller enclosed bays, Valun Bay provides sufficient natural barriers to maintain manageable conditions while allowing the influence of moderate currents. Its moderate size also facilitates efficient tracking and retrieval of drifters. Data collection was carried out using 12 TB-560 tracking beacons from Alltek Marine Electronics Corp (AMEC) \cite{ALTEK_Marine}, with all available drifters deployed to maximize measurement coverage. These beacons are simplified versions of Automatic Identification System (AIS) Class B units, offering a cost-effective solution for vessel tracking. AIS is widely used in maritime navigation to automatically share important information, including vessel identity, position, course, and velocity. Data exchange between vessels and shore stations improves maritime safety, aids in collision avoidance, and traffic management. Despite their basic design, these beacons efficiently transmit AIS data, making them suitable for small vessels, recreational boats, fishing boats, and buoy-based applications. AIS messages contain various categories of information essential for maritime communication and navigation, with standardized formats ensuring compatibility across vessels and shore-based stations. Table \ref{tab:ais_message_types} outlines the main types of AIS messages.

\begin{table*}[!htb]
	\footnotesize
	\centering
	\caption{Summary of AIS Message Types}
	\label{tab:ais_message_types}
	\begin{tabularx}{\linewidth}{Xrrl} 
		\textbf{Message} & Description & Example Use Cases \\ 
		\hline 
		Type 1  & Position Report Class A  & Real-time vessel position (Class A) \\ 
		Type 5  & Static and Voyage-Related Data & Vessel identification and voyage details \\ 
		Type 18 & Standard Class B Position Report & Real-time vessel position (Class B) \\ 
		Type 24 & Static Data & Essential vessel identification information \\ 
		Type 27 & Position Report for Long-Range AIS & Extended-range position updates \\ 
		Type 8  & Binary Broadcast & Custom data transmission for specific applications \\ 
		Type 22 & Channel Management & Channel management for AIS transmissions \\ 
		\\
	\end{tabularx}
\end{table*}

Outfitting floating buoys with AIS devices allows real-time tracking of sea surface velocity. Buoys transmit collected data at set intervals, offering authorities and rescue teams valuable insights into ocean currents and surface conditions. Once activated, the TB-560 usually acquires a GPS position within a minute and sends position reports using AIS message types 18 and 24.

\subsubsection{Drifter preparation and deployment}

\begin{figure}[htbp]
	\centering{\includegraphics[width=\linewidth]{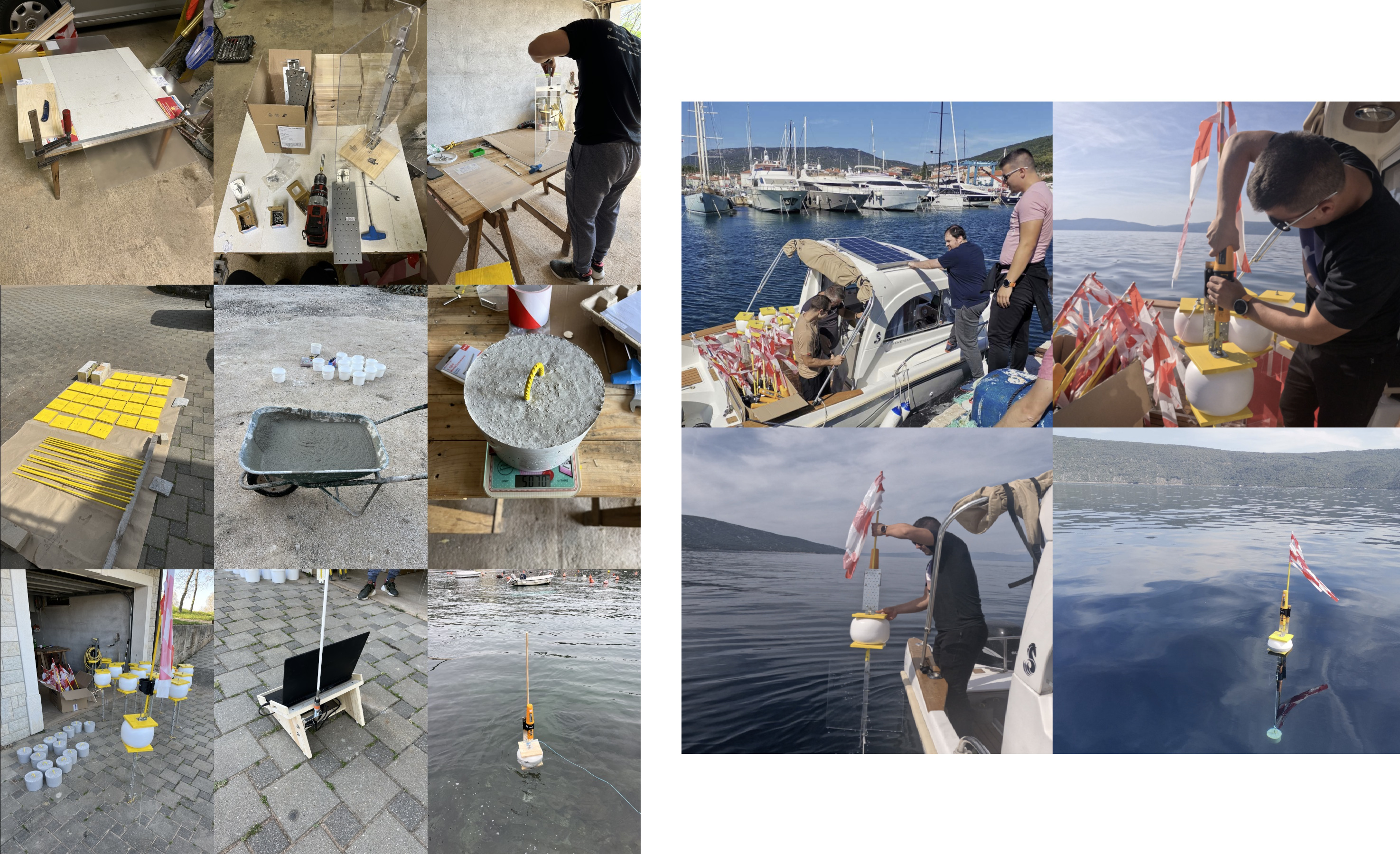}}
	\caption{Figure depicts the process of drifter preparation and deployment. On the left, the preparation process is shown, including the design and final assembly of the drifters. On the right, the loading of the drifters and their transit are illustrated. After reaching the deployment area, the drifters were activated and systematically deployed at pre-determined locations within Cres Bay to ensure optimal distribution for the experiment.}
	\label{fig:drifter_preparation_and_deployment}
\end{figure}

To measure sea surface velocity using AIS devices, a custom buoy platform with a mount for tracking beacons was designed and built. This involved developing a draught system to reduce the impact of waves and acquiring all necessary parts for constructing the drifters. A dedicated workstation, equipped with an AMEC Cypho-150 receiver and a 550 VHF antenna, was deployed along with the necessary software to decode and receive AIS messages from the tracking beacons. By combining these components, operational drifters were successfully built to capture sea surface velocity data. The entire process, from initial design to final construction, is shown on the left side of Fig. \ref{fig:drifter_preparation_and_deployment}. The right side of the figure depicts the loading of the equipment onto the boat and the subsequent deployment of all 12 available drifters during the experiment, which took place on September 21st and 22nd, 2024, to maximize our data collection.

\subsubsection{Experimental data}
Each drifter is assigned a distinct Maritime Mobile Service Identity (MMSI) number, which enables the retrieval of GPS coordinates linked to that MMSI from satellite systems. The drifter uses these GPS readings to calculate its speed and heading, transmitting the data every 10 seconds to a worksation through AIS messages 18 (position reports) and 24 (static data). During the data processing phase, these messages are decoded to extract the drifter's location and velocity. 

Since velocity measurements are taken every 10 seconds and rely on satellite GPS, some inaccuracies are unavoidable. To improve the accuracy of the data, we use a moving average filter on the raw measurements. Specifically, we compute the average velocity over 1-minute intervals, which involves averaging all the readings for each drifter within that period. This approach helps to minimize short-term variations, filtering out noise and highlighting more consistent trends. Averaging the data over a minute results in more accurate velocity measurements, reducing the impact of transient errors from individual data points. This method provides a more dependable depiction of surface velocity dynamics in our analysis.

\begin{figure}[htbp]
	\centering{\includegraphics[width=0.7\linewidth]{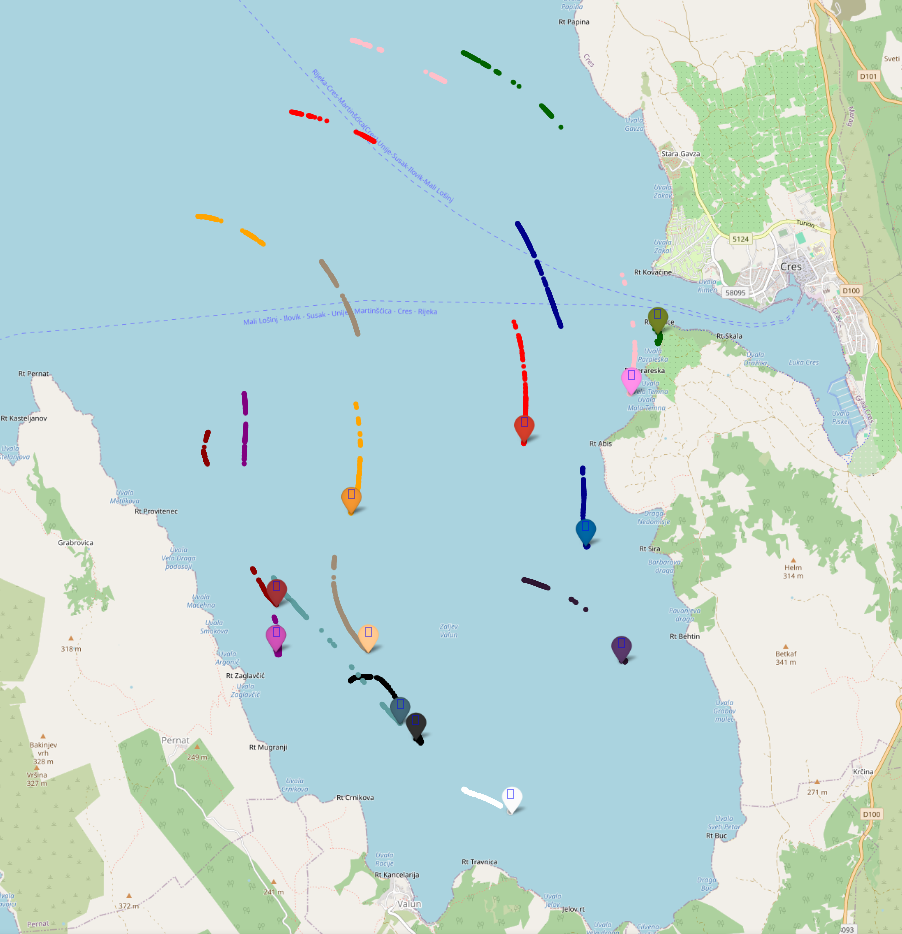}}
	\caption{Figure represents the trajectories of the drifters deployed during the sea experiment conducted at Valun Bay on September 22, 2024. Drifter movement was monitored over a four-and-a-half-hour period. The thick lines represent movement patterns influenced by local currents and environmental conditions.}
	\label{fig:Cres_trajectories}
\end{figure}

The drifter deployment across the 55-square-kilometer bay took approximately 1 hour and 30 minutes, with each drifter carefully positioned at its assigned location. The outcomes of the deployment on September 22, 2024, are shown in Fig. \ref{fig:Cres_trajectories}, with the movement of each drifter represented in different colors. 

While some measurement gaps are present, the overall movement pattern remains clear. These gaps are attributed to signal loss in specific areas of the bay, which may result from obstacles such as landforms, vegetation, or the distance between the drifters and the antenna. Despite these limitations, the deployment successfully produced valuable data on the drifters' movement and the dynamic conditions in the bay.

To replicate the trajectories of each drifter based on measurements, we focused on a 2-hour time period within the overall 4 hour and 30 minutes experiment. To capture drifter movement, we obtained new measurements at 300-second intervals, assuming that no drastic flow changes would occur within this time frame.

Initially, the algorithm with a fixed wind, as proposed in \cite{jakac2024approximation}, where a uniform wind field was added to the CFD simulation for flow reconstruction, was used. However, a key issue arose in the southern part of the domain, where actual velocities are smaller due to the coastline that surrounds the area, creating a closed system. The added wind velocity adversly affected the sea surface velocity in this region, causing the reconstructed trajectories to be significantly longer than those observed in the experiment.

\begin{figure}[htbp]
	\centering{\includegraphics[width=\linewidth]{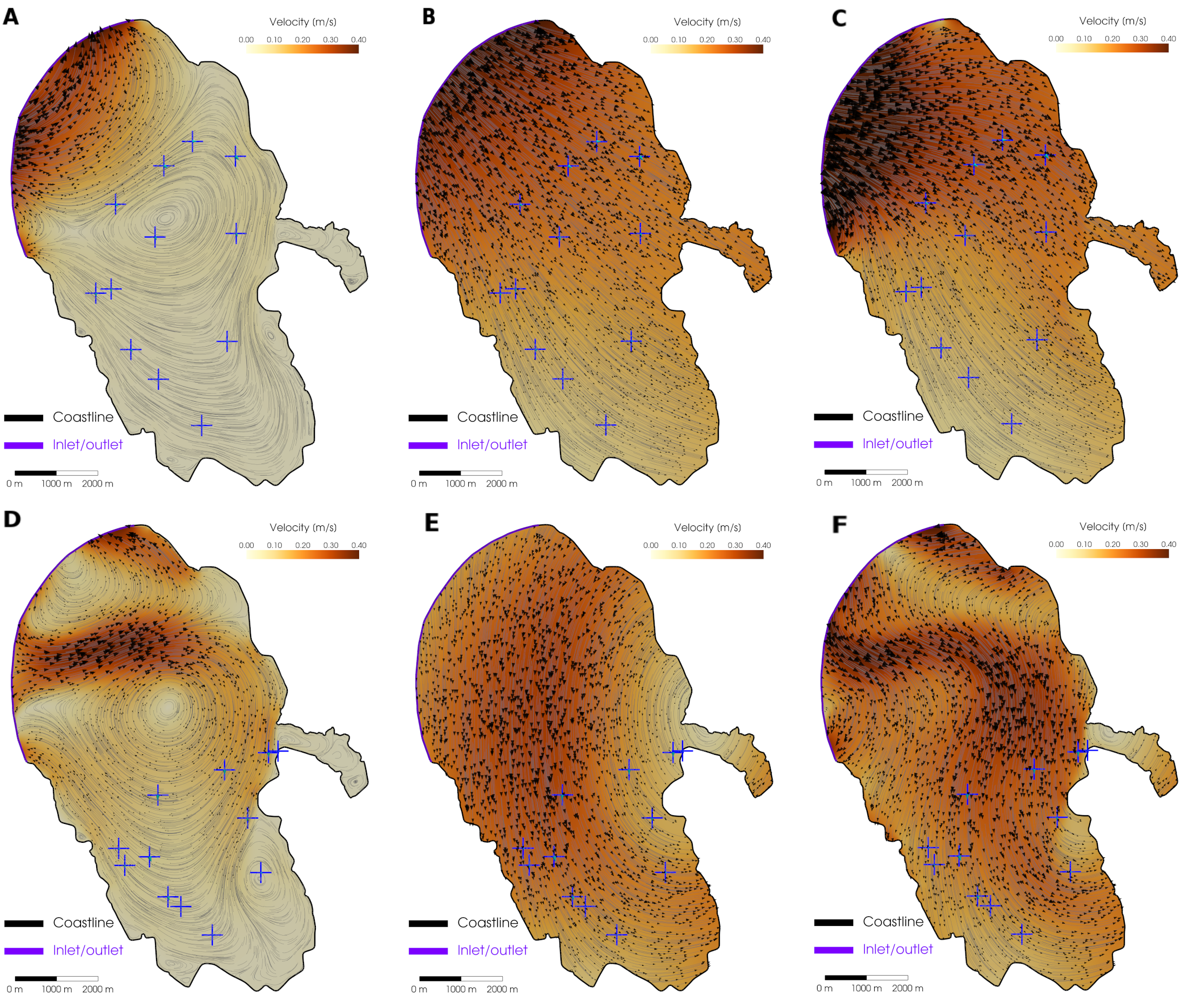}}
	\caption{Figure depicts the flow reconstruction using the fusion model approximation method at the initial and final time step. (\textbf{A}) The bounded domain CFD simulation at t = 0 s, with initial drifter locations marked by blue crosses. The simulation shows lower velocities in the southern part of the domain due to its shape, highlighting the limitation of CFD alone in capturing realistic movements in these areas.(\textbf{B}) Usage of open domain allows stronger velocities in the northern region and weaker velocities in the southern region, because of non-uniform open domain flow. (\textbf{C}) Reconstructed flow at t = 0 s, obtained by combining the bounded domain simulation with the open domain flow approximation. The resulting velocity field better captures surface dynamics due to the application of a fusion model. (\textbf{D}) The bounded domain CFD simulation of the domain at t = 7200 s shows significant changes from t = 0 s, as the drifters have moved over time, altering their direction and providing a realistic reference for the time-dependent simulation. (\textbf{E}) Open domain approximation at t = 7200 s, where the open domain flow has been adjusted to match the drifter trajectories, demonstrating the adaptability of the approximation model. (\textbf{F}) A notable change in the reconstructed flow compared to the initial (\textbf{C}) and final (\textbf{F}) time steps, highlighting the impact of fusion forcing on the velocity field. The alignment of the reconstructed flow with drifter movement indicates that the approximation effectively captures surface current variations over time.}
	\label{fig:cres_optimization_frames}
\end{figure}

To address this issue and accurately predict the movement of drifters within the closed system, a fusion model was employed with a updating measurement time step of $t_{M}$ = 300 seconds, based on data from the deployed drifters. At each $t_{M}$ new measurements from the deployed drifters were acquired. These measurements were then used to reconstruct the surface velocity field using a steady-state solver, with optimization running for $t_{M}$ seconds before new measurements arrived. To replicate the drifter movement, the positions of the drifters were advected over the $t_{M}$ time step using the reconstructed flow, ensuring that the flow dynamics remained consistent with the updated measurements at each step. Visual representation of changes at first and last time step can be seen on Fig. \ref{fig:cres_optimization_frames}

Based on results, conventional bounded domain simulation approach struggles to generate velocity fields that align with realistic measurements, particularly in regions far from the inlet and outlet, such as in closed domains like the Cres case. In contrast, the open domain approximation introduces non-uniform velocity fields and adjusts the overall flow to better match the measurements. The time-dependent evolution further emphasizes the model’s adaptability, improving its ability to represent realistic surface dynamics.

\begin{figure}[htbp]
	\centering{\includegraphics[width=\linewidth]{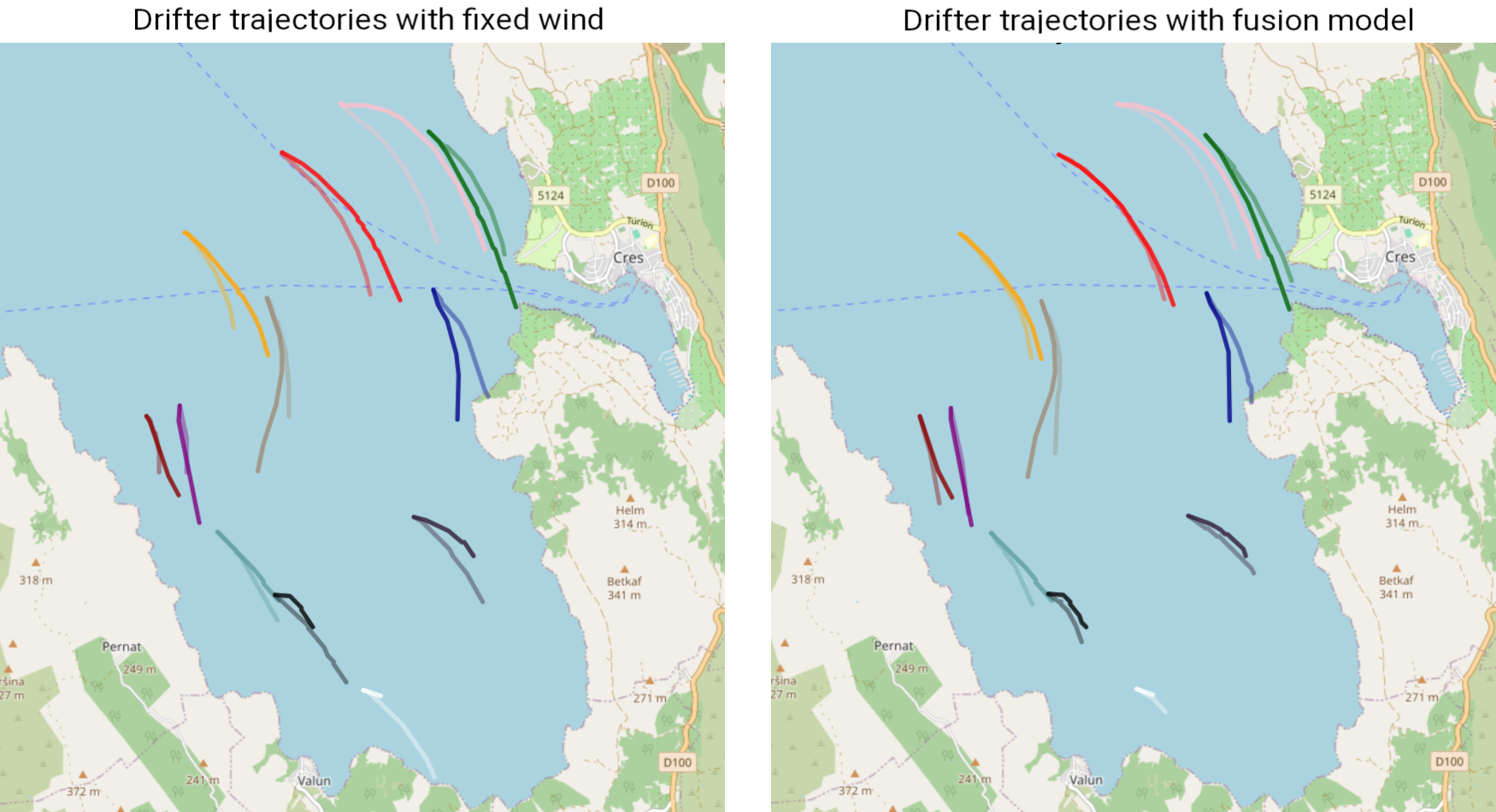}}
	\caption{The figure illustrates the difference in reconstructed experimental drifter trajectories using fixed and fusion model, with new measurements taken every 300 seconds over a 2-hour period. Darker colors represent the experimental drifter trajectories, while the reconstructed trajectories are displayed in the same color with increased transparency. It is evident that the non-uniform open domain flow better captures the realistic curvature of the trajectories, with several of the reconstructed trajectories showing high similarity to the experimental ones. }
	\label{fig:cres_opt_results}
\end{figure}

To validate the proposed methodology, we attempted to reconstruct experimental drifter trajectories using both fixed wind from \cite{jakac2024approximation} and fusion model. The difference is clearly illustrated in Fig. \ref{fig:cres_opt_results}, where experimental drifter trajectories are shown in darker colors, while the simulated trajectories for both correction approximations are displayed in the same color but with increased transparency. 

\begin{figure}[htbp]
	\centering{\includegraphics[width=\linewidth]{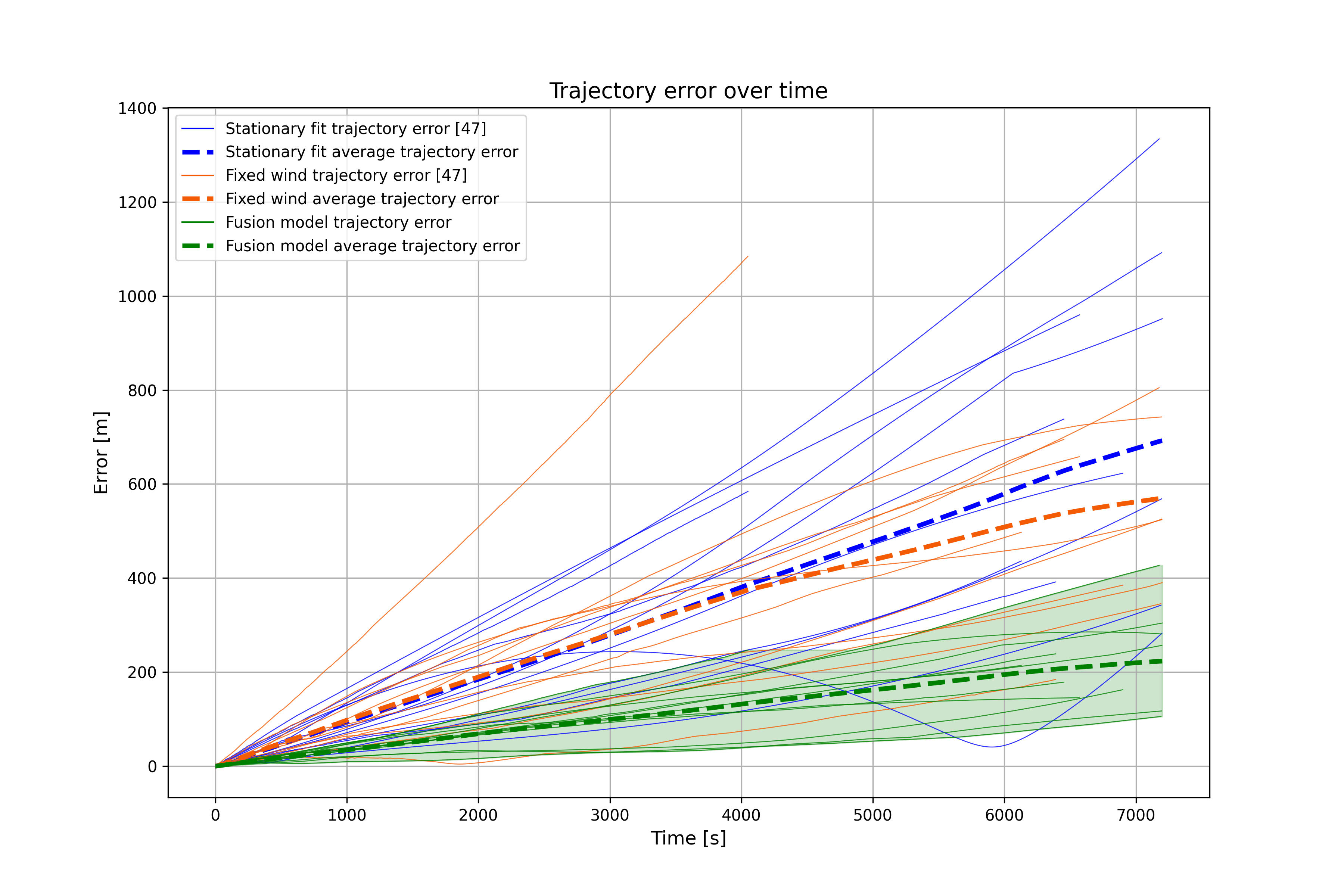}}
	\caption{The blue color on the graph represents trajectory reconstructions based on a stationary flow reconstruction with the fixed wind approximation, where the initial drifter locations are advected solely using the reconstructed flow from the initial time step. The orange color illustrates the approach using transient fit with fixed wind with updated measurements every 300 seconds in order to better guide optimization. This method shows improving results, particularly visible in the prolonged periods, as it adapts to new measurements, accounting for changes in the surface flow over time. The green color represents the use of the fusion model, combined with updated measurements every 300 seconds. The difference in trajectory reconstruction accuracy is evident, as the fusion model approximation results in a significant reduction in the average trajectory reconstruction error, offering a more precise match to the experimental data.}
	\label{fig:trajectory_comparison}
\end{figure}

The fusion model approximation provides notably improved results, more accurately capturing the realistic curvature of the trajectories, particularly in the southern part of the domain. In this region, the fixed wind approximation tends to exaggerate surface flow velocities, while the fusion model approximation effectively compensates for this issue. Although several of the reconstructed trajectories show a high degree of similarity to the experimental ones, some discrepancies are still present. To better assess the accuracy of the trajectory reconstruction, the trajectory reconstruction error, defined as the distance in meters between the reconstructed and experimental trajectories, is shown in Fig. \ref{fig:trajectory_comparison}.

It can be observed that both the stationary fit and fixed wind approaches exhibit large variations in trajectory areas, with the minimum and maximum error trajectories spanning a broad range over time. In contrast, the proposed methodology using the fusion approach demonstrates significantly smaller error trajectory areas, indicating a more consistent and replicable trajectory pattern.

\section{Limitations and discussion}

The proposed fused methodology combines two simplified two-dimensional flow models to act as a surrogate for submesoscale dynamics, aiming to predict the movement of a scalar through a stationary simulation with continuously updated measurements. Using coarse numerical meshes, as recommended, offers significant efficiency benefits, particularly for time-sensitive search and rescue operations. However, these meshes neglect smaller-scale and localized flows, as well as absolute accuracy, in favor of computational efficiency and global flow representation. Additionally, the model intentionally excludes factors like wind, tides, and temperature fluctuations to ensure computational efficiency, with these exclusions being addressed by the fusion model through an open domain flow approximation, thus balancing computational speed with predictive accuracy. While the proposed concept does not discourage the use of higher-quality meshes, it is important to be aware of the numerical consequences and the increased computational complexity that would result from their application.

Since the simplified two-dimensional steady flow model cannot fully replicate transient flows, we used an adaptive diffusion coefficient that adjusts based on the flow reconstruction error, accounting for inaccuracies in the reconstruction. While this improves the capture of the reference probability, it often leads to greater spreading of the scalar, meaning a larger area must be searched and potentially including regions that do not have the desired probability.

To validate our approach in a realistic scenario, we conducted a sea experiment using drifters to collect data, which was then used to reconstruct trajectories with updated measurements. One of the primary challenges in this process was the transmission rate, leading to gaps in the data, as well as potential GPS inaccuracies. These issues were partially mitigated by using interpolation to fill the gaps and applying a moving average approach to address GPS inaccuracies. Despite these difficulties, the method has been designed to align closely with experimental data, with the surrogate velocity field accurately replicating real flow patterns within an acceptable margin. A key feature of the approach is the updating of measurements at each measurement step, which enhances the optimization process while utilizing the simulation initialization method. This assumes minimal changes in the surface field between time steps, resulting in significantly faster computations, especially in more complex domains. However, this approach carries the risk of steering the optimization in the wrong direction, as all cases inherit the internal field from the best-found flow field. The optimization can still work effectively even without inheritance. However, introducing numerical complexity affects real-time applicability, reducing the window in which optimization can be performed, which is a key element influencing the accuracy of the reconstruction. Furthermore, when reconstructing experimental drifter trajectories, it becomes clear that the method may not fully capture all of the flow's movement, but it still provides very good results for trajectory approximations. Additionally, issues with field reconstruction arise when there are insufficient measurement points, making it difficult to assess field error and determine which optimization outcome aligns best with the results. This problem is more pronounced when measurement points are concentrated in a small area, limiting the accurate reconstruction of the flow across a larger domain. Therefore achieving a balance between a numerically simpler problem and a longer optimization window is very important.

\section{Conclusion}

Accurate sea surface velocity field approximation is crucial for applications like search and rescue operations or pollution spread monitoring, where the advection of passive scalars must be modeled in real-time or near real-time. However, achieving such accuracy is challenging even with advanced equipment like HF radars or complex oceanic flow simulations due to the dynamic nature of flows. Current methods, while effective in some aspects, often fail to provide detailed insights into the dynamic properties of flow fields and can often be computationally demanding, limiting their use in rapid response scenarios. 

In response, we present a data-driven framework that approximates the sea surface velocity field using scattered observation points. The framework utilizes a fused simplified two-dimensional surrogate flow model combined with an optimization approach to adjust boundary conditions, aligning the computed velocity field with measurements. By excluding factors like wind, tides, and temperature fluctuations, the method prioritizes speed, providing an efficient solution for real-world applications where quasi-steady flow assumptions are adequate for predicting advection in cases such as pollution dispersion and object tracking.

From a computational perspective, the use of coarse numerical meshes in our methodology improves efficiency, especially for time-critical applications. However, it's important to recognize that simplifying key environmental factors in the surrogate model may introduce inaccuracies in capturing the full complexity of real-case flow fields. These inaccuracies are acceptable when only an approximate velocity field is needed. Despite these simplifications, our approach offers significant advantages over traditional methods as it can mimic transient behavior and produce quasi-transient flow fields at specific time frames. This also ensures that the optimization can adapt to changing surface flow conditions, as demonstrated in the Cres sea experiment. Moreover, our method provides a reliable velocity field across the entire simulated domain, even in areas without nearby sampling points. This eliminates the need for extensive data, interpolation, or finite differencing, reducing both data collection time and costs. Overall, while challenges remain, the adaptability and efficiency of our methodology make it a valuable tool for real-time monitoring and decision-making in environmental applications.

\section*{Acknowledgments}
This publication is supported by the Croatian Science Foundation under the project UIP-2020-02-5090.

\section*{Data availability}
All parameters required to reproduce the study are provided in the manuscript. The data necessary for reproducing the presented method and cases can be accessed in The Open Science Framework repository: \url{https://osf.io/wjsb2/}. The Python code for reproducing this research is available upon request.


\bibliographystyle{elsarticle-num}  
\bibliography{references} 





\end{document}